\documentclass[pre,showpacs,preprintnumbers,amsmath,amssymb,superscriptaddress,twocolumn]{revtex4}

\usepackage{amsmath}
\usepackage{amssymb}
\usepackage{amsfonts}
\usepackage{fancyhdr}
\usepackage{fancybox}
\usepackage{ulem}

\usepackage{graphicx}
\usepackage{float}

\usepackage{amsthm}

\makeatletter

\newcommand{\OO}{{\cal O}}

\newcommand{\crl}[1]{[-\infty,\infty]}

\begin{document}
\title{Maxwell's Demon and Data Compression}
 \author{Akio Hosoya}
 \email{ahosoya@th.phys.titech.ac.jp}
 \affiliation{Department of Physics, Tokyo Institute of Technology, Tokyo, Japan}
 \author{Koji Maruyama}
 \email{maruyama@sci.osaka-cu.ac.jp}
 \affiliation{Department of Chemistry and Materials Science, Osaka City University, Osaka, Japan}
 \author{Yutaka Shikano}
 \email{shikano@th.phys.titech.ac.jp}
 \affiliation{Department of Physics, Tokyo Institute of Technology, Tokyo, Japan}
\date{\today}

\begin{abstract}
In an asymmetric Szilard engine model of Maxwell's demon, we show the equivalence between information theoretical and thermodynamic entropies when the demon erases information optimally. The work gain by the engine can be exactly canceled out by the work necessary to reset demon's memory after optimal data compression a la Shannon before the erasure.
\end{abstract}
\pacs{89.70.Cf, 05.70.-a}
\maketitle

\section{Introduction}
Entropy is one of the most cardinal concepts in the modern science. The idea of entropy plays a crucial role in not only thermodynamics, but also the physics of black holes and information science, including quantum information, to name a few. The interplay of entropy in classical physics and information science has been studied intensively since it was first pointed out by Brillouin in the general context~\cite{brillouin,Brillouin}. Then, this idea was clarified by Landauer in the form of information erasure principle~\cite{landauer}. Landauer's work opened up a way to relate the information theoretic and thermodynamic entropies. In order to obtain further insights into the relation, we need a simple and specific model. In this sense, the most well-studied is Szilard's engine and Maxwell's demon.

Since Maxwell mentioned an apparent violation of the second law of thermodynamics by a fictitious intelligent being in his textbook in 1871~\cite{maxwell}, this paradoxical problem has been debated intensively under the name of Maxwell's demon~\cite{demon2,maruyama09}. Towards its solution, Szilard devised in 1929 a one-molecule engine model, which ingeniously distilled the essence of the problem and made him realize the significance of information in the thermodynamic process~\cite{szilard}. Although it still took some time after Szilard, a satisfactory solution that lets the demon down was eventually reached, based on the idea of Landauer~\cite{landauer} and Bennett~\cite{bennett82}. The overall consensus we share today is that erasing information in demon's memory causes an entropy increase, which, with demon's best effort, precisely cancels out the work gain when closing the thermodynamic cycle.

The physical process of information erasure has been investigated from various aspects: noteworthy examples are two derivations of the entropy increase by Shizume~\cite{shizume95} and Piechocinska~\cite{piechocinska00}. They both showed that the lower bound of the entropy increase for erasing one bit of information should be $k_{B} \ln 2$, where $k_B$ is the Boltzmann constant. This entropy increase is exactly the minimum amount to circumvent the contradiction with the second law in the demonic paradox. 

This specific example suggests a possible way to link a certain entropy like quantity with the information entropy. This could be achieved by considering an operational model to carry out information erasure with a dynamical process intrinsic to the system of interest.

In the present paper, on the basis of the Shannon compression of demon's memory before erasure in the asymmetric Szilard engine model, we prove that the optimal cost of information erasure is
\begin{equation}
k_{B} \ln 2 \cdot H(p), \label{optcos}
\end{equation}
where $H(p)$ is the Shannon information entropy
\begin{equation}
H(p)= -p \log_2 p - (1-p) \log_2 (1-p)
\end{equation}
with $p$ being the ratio of the proportional division of the cylinder by the partition. Therefore, the entropy decrease of the Szilard engine exactly cancels out the entropy increase by the optimal information erasure of the demon.

This paper is organized as follows. In Sec.~\ref{sec:2}, we recapitulate the resolution of the Maxwell's demon paradox by Landauer and Bennett in the standard symmetric Szilard engine model. In Sec.~\ref{sec:3}, we propose the protocol of the demon in the case of an asymmetric Szilard engine. We show that the erasure work can be minimized by Shannon's data compression. In Sec.~\ref{sec:4}, we consider a different scenario, where heat baths of different temperatures are used for the engine-demon system. Section~\ref{sec:5} is devoted to summary and discussions.
\section{Symmetric Szilard engine and erasure of memory} \label{sec:2}
In this section we briefly review the Landauer principle of information erasure in the standard symmetric Szilard engine model~\cite{landauer}. The Szilard engine consists of a one-dimensional cylinder, whose volume is $V_0$, containing a single-molecule gas and a partition that works as a movable piston. 

The operator, i.e., a demon, of the engine inserts the partition into the cylinder, measures the position of the molecule, and connects to the partition a string with a weight at its end. These actions by the demon are optimally performed without energy consumption~\cite{bennett82}. Throughout this paper, the demon's memory is also modeled as a single-molecule gas in a box with a partition in the middle. Binary information, $0$ and $1$, is represented by the position of the molecule in the box, the left and the right, respectively. This model of symmetric memory has an advantage that reading, encoding, and computing over bits require no energy, making it consistent with the scenario of reversible computation.

\begin{figure}[ht]
\begin{center}
\includegraphics[width=7.5cm]{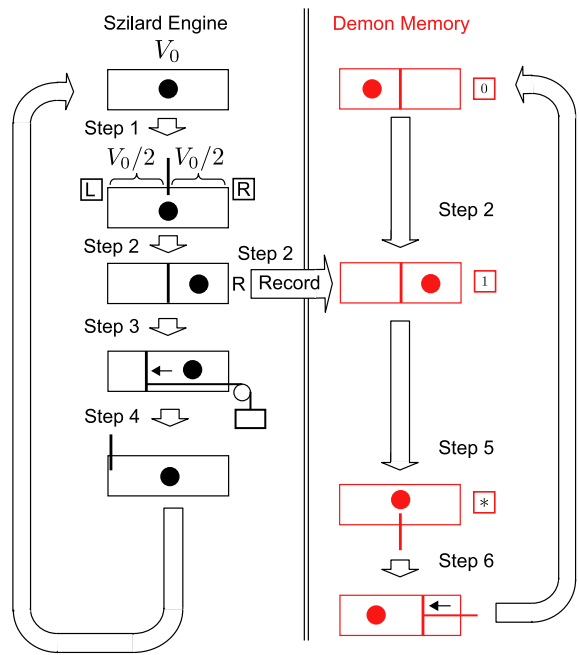}
\caption{(Color online). A protocol of symmetric Szilard engine (black/left side) and demon's memory (red/right side). This figure shows an example in which the molecule was found in the right hand side of the cylinder. In demon's memory, the state after removing the partition is denoted by ``$\ast$".} \label{fig:symmetric}
\end{center}
\end{figure}

The following is the protocol to extract work from the engine by information processing of the demon (see Fig.~\ref{fig:symmetric}), where we denote ``SzE" for the Szilard engine and ``DM" for the demon's memory at each step of the protocol. Initially, the molecule in the cylinder moves freely over the volume $V_0$. 
\begin{description}
\item[Step 1 (SzE)]
The partition is inserted at the center of the cylinder.
\item[Step 2 (SzE, DM)]
The demon measures the location of the molecule, either the left (``L") or the right (``R") side of the partition. The demon records the measurement outcome in his memory. When it is L (R), his memory is recorded as ``$0$" (``$1$").
\item[Step 3 (SzE)]
Depending on the measurement outcome, the demon arranges the device differently. That is, when the molecule was found on the left (right) hand side, i.e., the record is $0$ ($1$), he attaches the string to the partition from the left (right). In either case, by putting the cylinder in contact with the heat bath of temperature $T$, the molecule pushes the partition, thus exerting work on the weight, until the partition reaches the end of the cylinder. The amount of work extracted by the engine is
\begin{equation} 
	W= k_B T \ln 2, \label{eq:sz}
\end{equation}
as can be seen by applying the combined gas law in one dimension. 
\item[Step 4 (SzE)]
The demon removes the partition of the  engine, letting the molecule return to its initial state. 
\item[Step 5 (DM)] 
The demon removes the partition of his memory to erase information.
\item[Step 6 (DM)]
In order to reset the memory to its initial state, the demon compresses the volume of the gas by half.
\end{description}

In order to complete the cycle for both the Szilard engine and the memory, the demon has to reset the memory, which follows the erasure of one-bit information. Following is a more precise explanation about the physical process of information erasure and memory resetting described in Steps 5 and 6. The box is in contact with the thermal bath at the same temperature $T$ as that of the engine. The record in the memory can be erased simply by removing the partition, since the location of the molecule becomes completely uncertain. To bring the memory back to its initial state, e.g., $0$, one has to compress the gas by half by sliding a piston from the right end to the middle. The necessary work for this compression is $k_B T \ln 2$, which exactly cancels out the work gain by the engine (\ref{eq:sz}). Here, we have taken the result by Piechocinska for granted that the erasure of a single bit of information requires a work of at least $k_B T\ln 2$~\cite{piechocinska00}. 

Let us look at the same process in terms of thermodynamic entropy. By Steps 1 and 2, the volume of the gas in engine is halved, regardless of the measurement outcome. As the entropy change of an ideal gas under the isothermal process is given by $\Delta S := S ( V^\prime ) - S ( V ) = k_B \ln ( V^\prime / V )$, the entropy of the engine is lowered by $k_B \ln 2$. The isothermal expansion in Step 3 increases the entropy of the gas by $k_B \ln 2$, while that of the heat bath is decreased by the same amount. As far as the Szilard engine and its heat bath are concerned, the net result is an entropy decrease of $k_B \ln 2$. Nevertheless, this is exactly canceled out by the entropy increase due to information erasure and reset performed in Steps 5 and 6.

These last two steps are of crucial importance when closing a cycle of the memory. Information erasure in Step 5 is an irreversible process and increases thermodynamic entropy by $k_B \ln 2$. The isothermal compression to reset the memory in Step 6 requires work and dissipates entropy of $k_B \ln 2$ to its heat bath. This is the essence of Landauer-Bennett mechanism that resolves the Maxwell's demon paradox.

\begin{figure}[ht]
\begin{center}
\includegraphics[width=5cm]{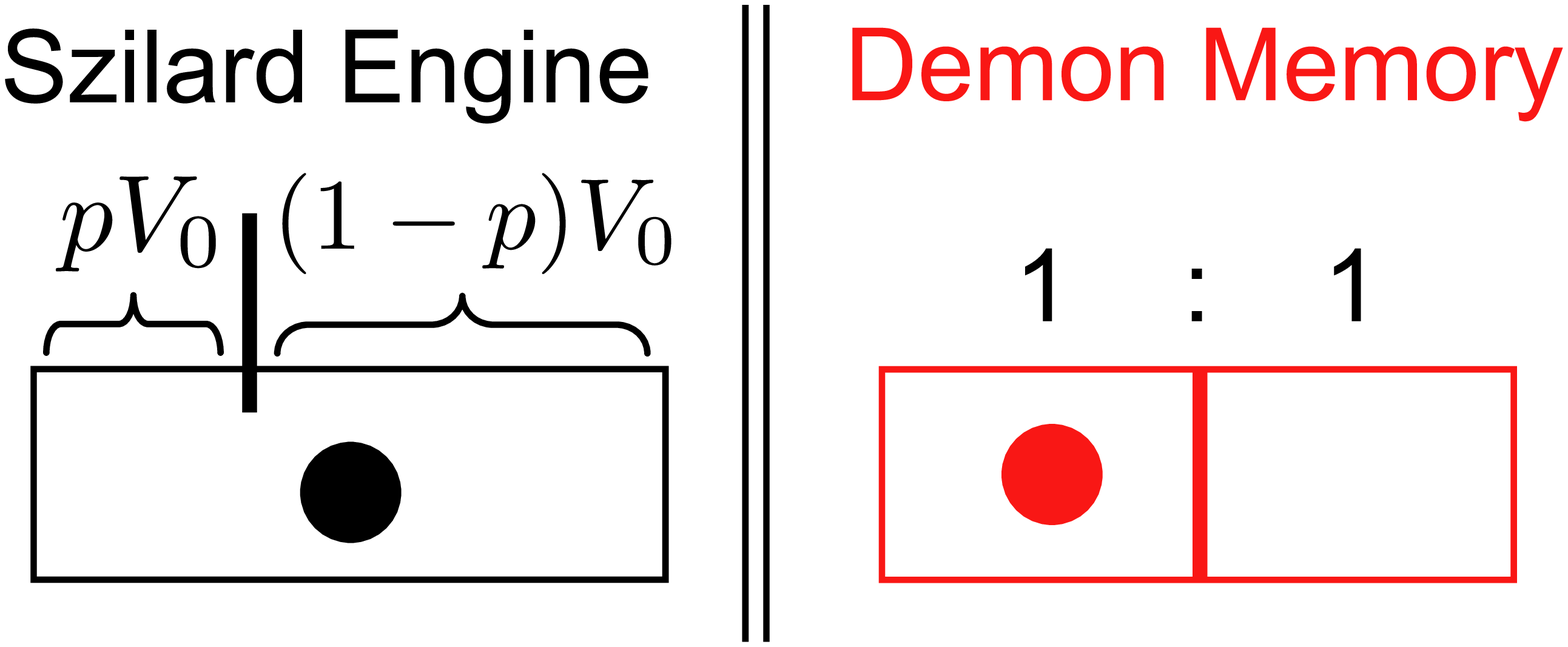}
\caption{(Color online). The model of an asymmetric Szilard engine. The position of the partition in demon's memory is the same as that in the case of the symmetric version.} \label{fig:asymmetric}
\end{center}
\end{figure}

Now let us slightly generalize the Szilard engine model to an asymmetric one in such a way that the partition is inserted to divide the whole volume 
$V_0$ into $p V_0$ and $(1-p) V_0$ with $0 < p < 1$ (See Fig.~\ref{fig:asymmetric}). A straightforward calculation shows that the work extracted 
by the asymmetric Szilard engine is
\begin{equation}
	k_B T S(p), \label{shaeq}
\end{equation}
where $S(p) = - p \ln p - (1-p) \ln (1-p)$~\cite{Feynman}. If the memory is reset after every cycle of the engine, the amount of work consumption is 
\begin{equation}
	\Delta W = k_B T \ln 2 - k_B T S(p) \geq 0.
\end{equation}
However, one may wonder if the gap, $\Delta W$, could be smaller by employing a better strategy. In the following section, we show an information theoretical protocol that fills the gap optimally.
\section{Asymmetric Szilard engine and erasure of compressed memory} \label{sec:3}
We are going to show a protocol in which the demon is clever enough to reduce the work for the erasure by using the Shannon data compression~\cite{shannon} in the asymmetric Szilard engine introduced in the previous section. First, the demon accumulates the data of $N$ cycles, which we assume is very large. The data contains uneven number of $0$'s and $1$'s corresponding to the measured position of the molecule in the engine. The relative frequency of $0$'s is obviously $p$, while that of $1$'s is $1-p$. According to Shannon's noiseless coding theorem, the demon can compress the data to a shorter one, whose length will be $N_s := N H(p) \leq N$ at shortest. Coding does not cost any work if we employ reversible computation~\cite{bennett_rev}, provided that the memory is symmetric as remarked before. If asymmetric memory were used, even the NOT gate, and therefore generic computation, cost energy, which makes our task less transparent. See, e.g., Refs.~\cite{barkeshli,sagawa}. Then, he erases the shortened data string with the work $k_B T \ln2 \cdot N_s = k_B T N S(p)$. Therefore, the difference between the work to reset the memory and the work extracted by the engine approaches zero,
\begin{equation}
	\Delta W(optimal)= k_B T N_s - k_B T S(p) = 0,
\end{equation}
for a very large $N$.

To be more precise, we write down the optimal protocol below.
\begin{description}
\item[Step 1 (SzE)]
The partition is inserted to divide the volume into two parts, $pV_0$ and $(1-p)V_0$, in the initial configuration of the cylinder and a single molecule is either on the left or the right of the partition.
\item[Step 2 (SzE, DM)]
The demon measures the location of the molecule and records either $0$ for the left (L) or $1$ for the right (R) and keep the result in his memory.
\item[Step 3 (SzE)]
Depending on the recorded information, the demon arranges the device differently. That is, when the molecule was found on the left (right) hand side, i.e., the record is $0$ ($1$), he attaches the end of the string to the partition from the left (right). In either case, the molecule pushes the partition which is now movable to the very end of the cylinder.
\item[Step 4 (SzE)]
In order to go back to the initial configuration, the demon disconnects the cylinder from the attached device.
\item[Step 5 (DM)]
The demon repeats Steps from 1 to 4 for $N$ times, keeping the $N$-bit string in his memory. Then, he compresses the $N$-bit string to the minimum length $N H(p)$, according to Shannon's noiseless coding theorem. We break up the $N=mn$ bit string into $m$ blocks of $n$ bits.

A brief description of the data compression is the following.

First, punctuate the $N$-bit string by $n$ bits so that we have $2^n$ sequences with relative frequencies $p^n, (1-p)p^{n-1}, \dots, (1-p)^{n}$ so that we can encode the sequences into strings of bit length of $-\log_2 p^n, -\log_2 (1-p)p^{n-1}, \dots, -\log_2 (1-p)^{n}$, if they were integers. Roughly speaking, the average bit length would be
\begin{align}
	& \sim \sum^{n}_{k=0} {n \choose k} \left\{ - (1-p)^{k}p^{n-k} \log_2 (1-p)^{k}p^{n-k} \right\} \notag \\ 
	& = n H(p). \label{rough_data}
\end{align}
The right hand side of Eq. (\ref{rough_data}) coincides with the shortest average bit length shown by Shannon~\cite{shannon}.
\item[Step 6 (DM)]
The demon repeats the process of $n$ turns $m$ times so that the total amount of bits to be erased is
\begin{equation}
	\tilde{S} \approx m n H(p) = N H(p). \label{shannon_comp_eq}
\end{equation}
\end{description}
\begin{figure*}[ht]
\begin{center}
\includegraphics[width=12cm]{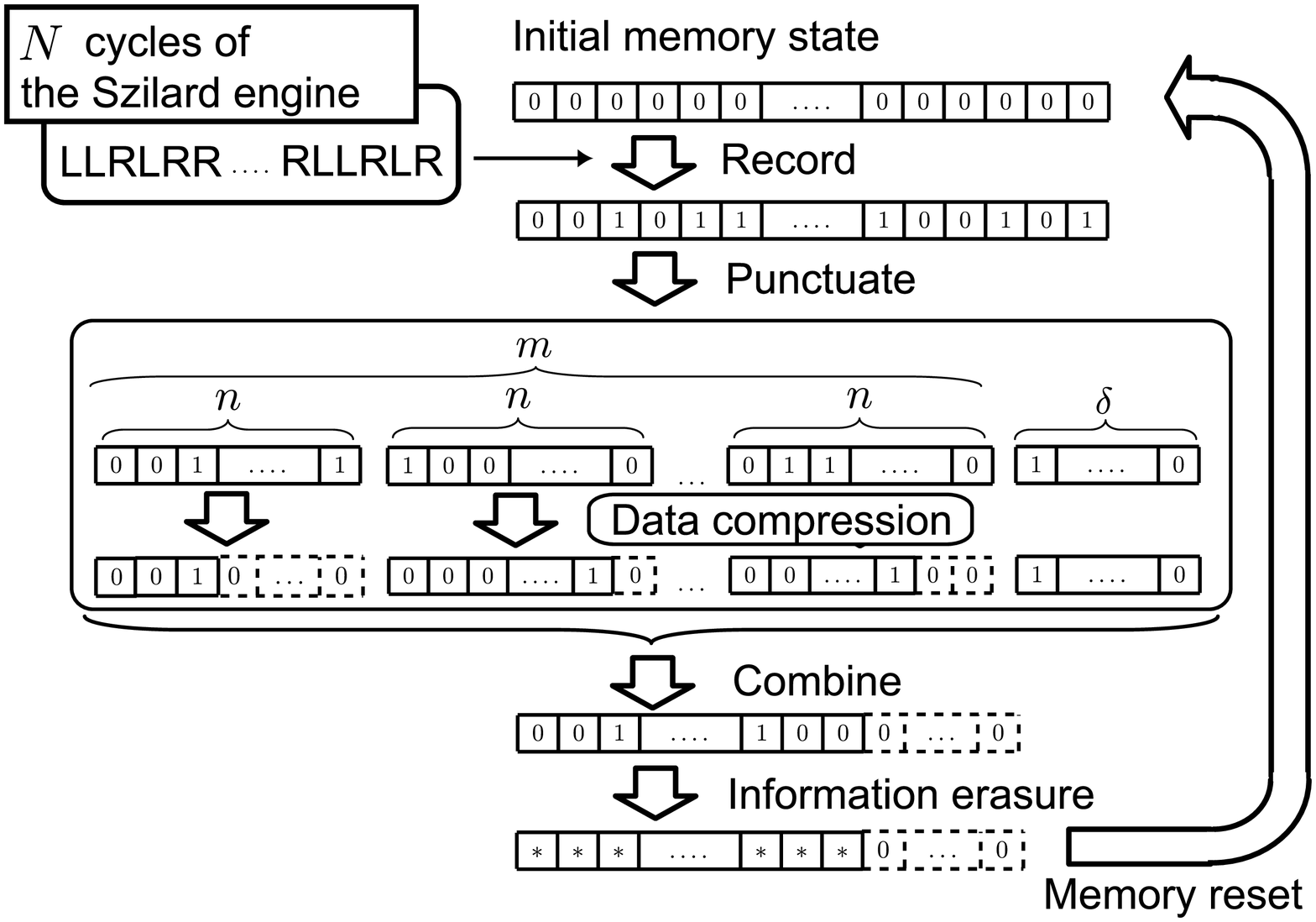}
\caption{The protocol of data compression for demon's memory. The dashed blocks express the trivial initial memory state ``0" after data compression.} \label{datacomp}
\end{center}
\end{figure*}

The change in thermodynamic entropy is calculated in the same manner as in the above case of symmetric engine. The volume of the gas after Step 2 becomes $p V_0$ with probability $p$ or $(1-p)V_0$ with probability $1-p$. The entropy of the gas after Step 2 is thus decreased by $- p \ln p - (1-p) \ln (1-p)$, which is equal to $S(p)$ in Eq. (\ref{shaeq}) and is to be canceled out by the later steps of information erasure and memory resetting.

Let us treat Steps 5 and 6 more precisely. According to the Shannon source coding theorem for the symbol codes~\cite{shannon}, for $n$-bit string, there always exists an {\it optimal code} such that the averaged code length $\bar{S}$ satisfies $n H(p) \leq \bar{S} < n H(p) + 1$. The well-known example of optimal codes is the Huffman code for the encoding procedure, see Ref.~\cite{huffman}. It is reminded that the demon breaks up the $N$-bit string into $n$-bit block, i.e., $N = n m + \delta$, where $0 \leq \delta < n$. The extra $\delta$-bit string cannot be encoded. When the demon optimally encodes an $N$-bit string, its length is longer than $N H(p)$ bits by $m + \delta$ bits at worst. As the discrepancy is bounded as 
\begin{equation}
	m + \delta = N - (n - 1) m \geq N - \frac{(n-1)^2 + m^2}{2},
\end{equation}
it can be minimized when $n - 1 = m$. Thus, we obtain $n = m = \OO (\sqrt{N})$. The extra bits are at worst $\delta = \OO (\sqrt{N})$. Therefore, we can more precisely express the average length of the optimally compressed data string as 
\begin{equation}
	\tilde{S} = m n H (p) + \delta = N H (p) + \OO (\sqrt{N}).
\end{equation}
The averaged work necessary to erase information in this string over $N$ bits is 
\begin{align}
	W (erasure) & = \frac{k_B T \ln 2 \cdot \tilde{S}}{N} \notag \\ 
	& = k_B T \ln 2 \cdot H (p) + \OO \left( \frac{1}{\sqrt{N}} \right).
	\label{opt_era}
\end{align}

On the other hand, in Step 3, the amount of average work extracted by the engine over $N$ cycles is given by 
\begin{equation}
	W (engine) = k_B T S (p) + \OO \left( \frac{1}{\sqrt{N}} \right),
	\label{opt_work}
\end{equation}
where $S(p) = \ln 2 \cdot H (p)$ as can be seen by applying the combined gas law. It is reminded that $S(p)$ is the thermodynamic entropy as discussed before.

It is now clear that the work for information erasure of demons's memory (\ref{opt_era}) and the work from the asymmetric Szilard engine (\ref{opt_work}) agree for sufficiently large $N$. The above argument leads us to conclude that the information theoretical entropy is equivalent to the thermodynamic entropy when optimal information processing is physically executed.

Note that, for the symmetric Szilard engine, we do not need to compress data because the number of $0$'s and $1$'s are equal. Also, the erasure model for non-equiprobability distribution of the memory was considered in a simple thermodynamic process~\cite{maruyama09,maruyama_phd}. The amount of work for this process coincides with the optimal one~(\ref{opt_era}).
\section{Another heat bath at a lower temperature} \label{sec:4}
\begin{figure}[ht]
\begin{center}
\includegraphics[width=8cm]{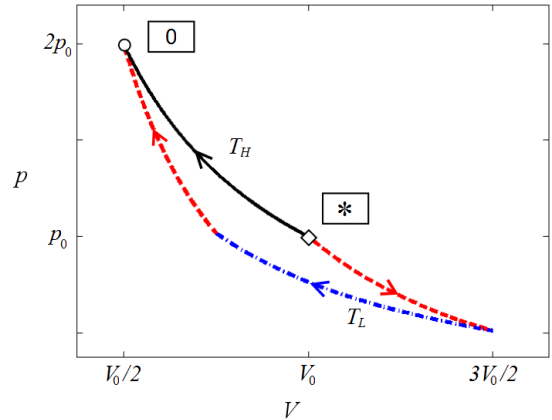}
\caption{(Color online). The $p$-$V$ diagram of an alternative erasing process with a heat bath of lower temperature. The erasure is realized with any path from the ``$\ast$" state to the ``$0$" state, which are denoted by a diamond $\diamond$ and a circle $\circ$, respectively. The solid black line represents the standard erasure by an isothermal compression (at temperature $T_H$). Although the demon may want to use a colder heat bath of temperature $T_L$, the entropy increase due to the erasure cannot be smaller than $k_B \ln 2$. The path, consisting of two adiabatic processes (red dashed lines) and one isothermal compression (blue dot-dashed line), in the figure is the optimal one in terms of entropy increase and attains the Landauer limit of $k_B \ln 2$. Naturally, the entropy cost is the same even if the temperature of the memory is always $T_L$, while the Szilard engine is operated at $T_H$.} \label{diagram}
\end{center}
\end{figure}

The reader might question why the demon resets his memory at the same temperature (say, $T_H$) as the heat bath for the Szilard engine and wonder what if the erasure is executed at a lower temperature, $T_L (<T_H)$, because the compression of the memory space would then require less work. With two heat baths of different temperatures, some nonzero work $W$ can indeed be extracted, however, the amount of entropy increase is always larger than or equal to $k_B \ln 2$. That is, in terms of entropy balance there is no difference from the case with a single heat bath. Hence, the demon's attempt to outdo the second law ends up in vain, as we naturally expect. An example of erasing process with two heat baths that achieves the bound is depicted in Fig. \ref{diagram} and explained in its caption.

When the optimality is achieved, the entire compound system, the engine and the memory, simply works as a single engine; it converts a part of heat absorbed at $T_H$ into the work $W$ and throws the residual energy away to the heat bath at $T_L$. The overall thermal efficiency is equal to $\eta= W / Q_H = 1 - T_L / T_H$, where $Q_H$ is the amount of heat flowed from the heat bath at $T_H$ to the Szilard engine, thus effectively the same as the Carnot engine.

Let us make a remark to avoid a potential confusion. Despite being equivalent to the Carnot engine, the engine-memory system does not work reversibly in the context of information erasure. While the system can be run in the reverse direction, the erased information can never be restored reliably.
\section{Summary and Discussions} \label{sec:5}
We have shown in the asymmetric Szilard engine that the work extracted by Maxwell's demon is asymptotically canceled out after a large number of cycles by the work to reset the memory after optimal data compression. We have described an explicit protocol and shown its optimality by making use of Shannon's noiseless coding theorem. The key point is data compression before information erasure of the memory and this argument makes the seminal work by Landauer and Bennett more general and precise. The coincidence between information and thermodynamic entropies is now very clear, thanks to the demon's cleverest strategy. As a slight generalization we have also considered the case of information erasure at lower temperature to see that the efficiency of the whole system can be only as efficient as the Carnot cycle and that there is no net gain for the demon.

We would like to stress that the thermo-informational cycle has to be completed to correctly address the apparent violation of the second law in the context of Maxwell's demon. This means that no residual information should be left outside the engine-demon system after the cycle \cite{bennett03}. What makes the argument of demon important and interesting is this physical loss of information, otherwise it is merely a sequence of normal measurements.

As briefly remarked in the introduction, the present work in the specific model suggests a general method to relate information and physical entropies by considering an operational process to erase information in the physical system. One such example is the original derivation of the black hole entropy by Beckenstein~\cite{beckenstein}. He considered a gedanken experiment of dropping a ``particle of one bit" for information erasure which increases the area of the event horizon as a back action. He identified the amount of information loss with the change of the intrinsic entropy of black hole.  Also, there is a well-known derivation of the Boltzmann distribution on the basis of the principle of the maximum Shannon entropy under the energy constraint~\cite{Jaynes}. However, the physical meaning of the Shannon entropy there is not clear, though the optimal value coincides with the thermodynamic entropy. It would be nice if we could clarify the meaning of the maximization of the Shannon entropy in terms of the optimal memory reset.

The optimal information erasure would help us fully understand physical entropy in terms of information entropy, as Brillouin envisioned~\cite{Brillouin}.
\section*{Acknowledgments}
The authors would like to thank Haruka Kibe for her contribution in the early stage of the present investigation and Charles Bennett for useful discussion. The authors (A.H. and Y.S.) are supported by the Global Center of Excellence Program ``Nanoscience and Quantum Physics" at Tokyo Institute of Technology. K.M. is supported by Grant-in-Aid for Scientific Research (C) (No. 22540405). Y.S. is also supported by JSPS (Grant No. 21008624).

\end{document}